\documentclass[12pt]{iopart}
\usepackage{amssymb}
\usepackage{inputenc}
\usepackage{graphicx}
\usepackage{units}
\usepackage{color}
\usepackage{braket}

\begin{document}

\title{Systematic study of tunable laser cooling for trapped-ion experiments}
\author{A. P. Kulosa$^1$, O. N. Prudnikov$^{3,4}$, D. Vadlejch$^1$, H. A. F\"{u}rst$^{1,2}$, A. A. Kirpichnikova$^3$, A. V. Taichenachev$^{3,4}$, V. I. Yudin$^{3,4,5}$, and T.E. Mehlst\"{a}ubler$^{1,2}$}
\address{$^1$ Physikalisch-Technische Bundesanstalt, Bundesallee 100, 38116 Braunschweig, Germany}
\address{$^2$ Institut f\"{u}r Quantenoptik, Leibniz Universit\"{a}t Hannover, Welfengarten 1, 30167 Hannover, Germany}
\address{$^3$ Institute of Laser Physics, 630090, Novosibirsk, Russia}
\address{$^4$ Novosibirsk State University, 630090, Novosibirsk, Russia}
\address{$^5$ Novosibirsk State Technical University, 630073, Novosibirsk, Russia}
\ead{andre.kulosa@ptb.de}


\begin{abstract}
We report on a comparative analysis of quenched sideband cooling in trapped ions. We introduce a theoretical approach for time-efficient simulation of the temporal cooling characteristics and derive the optimal conditions providing fast laser cooling into the ion's motional ground state. The simulations were experimentally benchmarked with a single $^{172}$Yb$^+$ ion confined in a linear Paul trap. Sideband cooling was carried out on a narrow quadrupole transition, enhanced with an additional clear-out laser for controlling the effective linewidth of the cooling transition. Quench cooling was thus for the first time studied in the resolved sideband, intermediate and semi-classical regime. We discuss the non-thermal distribution of Fock states during laser cooling and reveal its impact on time dilation shifts in optical atomic clocks.  
\end{abstract}

%
\vspace{2pc}
\noindent{\it Keywords}: Tunable laser cooling, atom kinetics, cooling dynamics, trapped ions
%
%
%
%

\section{Introduction}
Laser cooling is an essential tool for modern quantum optics experiments with trapped ions, such as the study of topological defect formation in ion Coulomb crystals~\cite{Mielenz2013,Ulm2013,Ejtemaee2013,Kiethe2021}, quantum simulation~\cite{Blatt2012,Joshi2020a,Monroe2021} and quantum computation~\cite{Gill2022}. Highly-accurate ion optical atomic clocks are traditionally operated at the Doppler cooling limit~\cite{Huntemann2016,Huang2022}. As a consequence of their constantly improving frequency uncertainty, the time dilation due to the residual ion secular motion at Doppler temperature nowadays poses a limiting contribution to the clock’s error budget at the low $10^{-18}$ level. In order to realize ion optical clocks at even lower uncertainties, they will have to operate at ultra-cold temperatures~\cite{Brewer2019,Keller2019a}.   

Since the first implementation of laser cooling (see e.g.~\cite{Phillips1998}), several sub-Doppler cooling mechanisms have been developed and improved. Electromagnetically induced transparency (EIT) is an established tool to rapidly cool trapped ions into the motional ground state~\cite{Morigi2000,Roos2000}. Recently it was applied to cool the strong transverse mode in a single $^{171}$Yb$^+$ ion to the motional ground state within a few $\unit[100]{\mu s}$~\cite{Feng2020,Qiao2021}. Similar cooling rates were reported for cooling with polarization gradients (PG)~\cite{Dalibard1989} created by a standing wave field along the trap axis~\cite{Joshi2020}. PG cooling of all motional modes in a single $^{171}$Yb$^+$ ion to $\bar{n}\simeq 1$ was so far achieved within a few \unit[]{ms}~\cite{Ejtemaee2017}.

Besides the aforementioned schemes, sideband cooling in the Lamb-Dicke regime is a well established technique for trapped ions~\cite{Wineland1979}. Cooling on a narrow optical transition can be further enhanced by optical quenching, i.e. a laser-induced increase of the natural linewidth $\gamma/2\pi$ of the cooling transition~\cite{Diedrich1989,Marzoli1994,Mehlstaeubler2003}. This method has been used in both trapped ion~\cite{Letchumanan2007,Sawamura2008,Chiaverini2014} and neutral atom experiments~\cite{Binnewies2001,Rehbein2007}. The effective linewidth $\gamma_{\rm eff}/2\pi$ is steered with the intensity of the quenching laser, which couples the excited state of the cooling transition to a short-lived intermediate state (see fig.~\ref{fig:quantummodel}a). This allows for laser cooling in a broad range of different regimes ranging from the resolved sideband regime ($\gamma_{\rm eff}\ll\omega_{\rm osc}$) to the semi-classical regime ($\gamma_{\rm eff}>\omega_{\rm osc}$), which can be described in terms of sub-Doppler~\cite{Dalibard1989,Joshi2020} or Doppler (see e.g.~\cite{Minogin1987,Metcalf1999,Wesenberg2007,Prudnikov2017}) forces acting from the resonant light field on the ion. Here, $\omega_{\rm osc}$ is the motional frequency the ion exhibits in the harmonic pseudopotential of the radio-frequency (rf) Paul trap. So far, optimisation studies of quench cooling were dedicated either to the limits of low saturation on the cooling transition~\cite{Marzoli1994,Javanainen1981,Javanainen1984} or to the strong-sideband regime, where $(\gamma_{\rm eff}/2\omega_{\rm osc})^2\ll\eta$~\cite{Blockley1993,Zhang2021}. $\eta=\sqrt{\omega_{\rm rec}/\omega_{\rm osc}}$ is the Lamb-Dicke parameter relating the ion's kinetic energy change due to photon recoil with frequency $\omega_{\rm rec}$ to its confinement in the harmonic potential.

In this work, we go beyond the aforementioned limitations by studying optimised cooling in the sideband, semi-classical and intermediate regime ($\gamma_{\rm eff}\simeq\omega_{\rm osc}$). We particularly study the impact of these regimes onto the distribution of atomic Fock states. The intermediate regime is naturally present for e.g. cooling of In$^+$ ions~\cite{Peik1999} and can be used to cool the motional modes of multiple ions in the radial directions simultaneously. We especially focus on the nonlinear dependence of cooling time on cooling field intensity. We present a versatile method for the calculation of the characteristic cooling time, which does not require solving the dynamical evolution of the system’s density matrix. With this, we determine the optimal conditions for fast and deep laser cooling with significantly reduced computational efforts when compared to the full density matrix approach. We derive general analytical expressions for the optimal cooling laser Rabi frequency and the minimum cooling time which can be applied to any trapped ion species. 

The simulation results are benchmarked against experimental data acquired with a single $^{172}$Yb$^+$ ion confined in a high-precision rf Paul trap~\cite{Keller2019}. We study the temporal evolution of the Fock state distribution during quench cooling and discuss its influence on the thermal time dilation shift in atomic clocks. Our findings pave the way towards fast and deep cooling of larger ion ensembles arranged ion Coulomb crystals. 

This paper is organized as follows: in \textbf{Chapter 2} we recall the quantum model used for ion-light interaction and its reduction to an effective two-level system in the frame of optical quenching. \textbf{Chapter 3} describes our simulation approach of using the ``${\hat \tau}$-matrix'' for faster computation of cooling dynamics, which is used for a systematic study of the sideband cooling regime in \textbf{Chapter 4}. In \textbf{Chapter 5} we apply our ``${\hat \tau}$-matrix method'' to the specific case of an $^{172}$Yb$^+$ ion confined in a rf Paul trap and study cooling ranging from the resolved sideband to the Doppler regime. We finally discuss the non-thermal distribution of Fock states during cooling and its impact on time dilation shifts in optical atomic clocks in \textbf{Chapter 6}.    


\section{Quantum model for ion-light interaction}
\begin{figure}[t]
    \centering
    \includegraphics[width=0.6\textwidth]{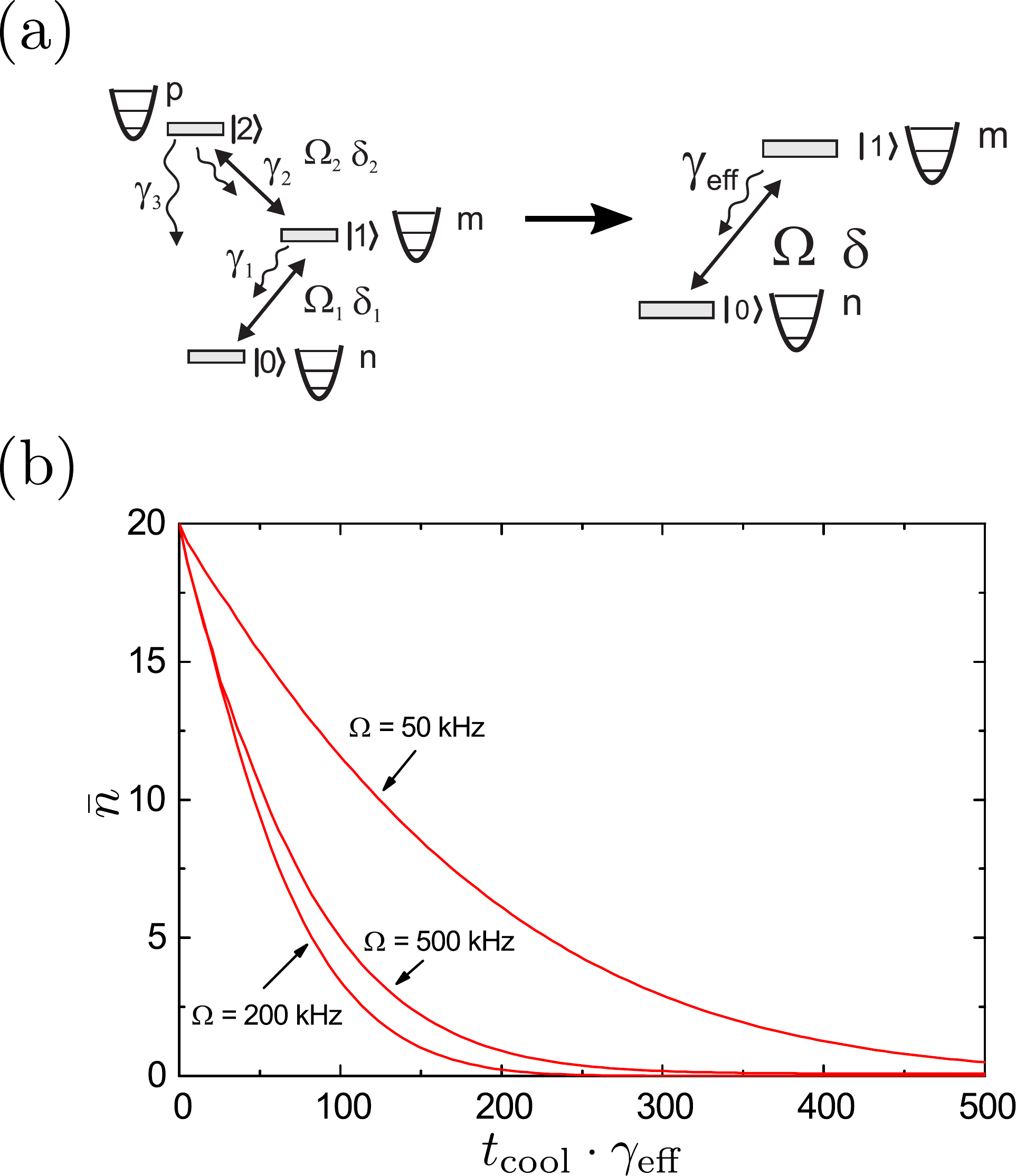}
    \caption{(a) In a three-level system, the decay rate $\gamma_1$ of state $\ket{1}$ can be increased by laser-coupling to a higher-lying state $\ket{2}$, which features a fast decay $\gamma_3\gg\gamma_1$ to the ground state $\ket{0}$. For low saturation on the $\ket{1}\to\ket{2}$ transition, state $\ket{2}$ can be adiabatically eliminated resulting in an effective two-level system with decay rate $\gamma_{\rm eff}$ of state $\ket{1}$~\cite{Marzoli1994}. (b) Temporal evolution of the mean occupation number during quench cooling with $\gamma_{\rm eff}/2\pi=\unit[50]{kHz}$ and $\bar{n}_{\rm ini}=20$. The secular frequency is set to $\omega_{\rm osc}/2\pi =\unit[600]{kHz}$ and the cooling light detuning is $\delta=-\omega_{\rm osc}$. The fastest cooling rate is observed for a specific Rabi frequency $\Omega_{\rm opt}$ of the cooling light, which is not expected from previous theoretical discussions~\cite{Javanainen1981}. }
    \label{fig:quantummodel}
\end{figure}

The cooling dynamics of an ion confined in a rf Paul trap can be described by the quantum kinetic equation for the density matrix in single particle approximation

\begin{equation}\label{dm}
\frac{\partial {\hat \rho}}{\partial t} = -\frac{\mathrm{i}}{\hbar}
\left[{\hat H}, {\hat \rho} \right] +{\hat \Gamma}\{ {\hat \rho}\},
\end{equation}
where ${\hat H}$ is the Hamiltonian and the term $\hat\Gamma\{\hat\rho\}$ describes the relaxation of the density matrix due to spontaneous decay. The
Hamiltonian is composed of ${\hat H}={\hat H}_{\rm ext}+{\hat H}_{\rm int}+{\hat
V}_1+{\hat V}_2$, where

\begin{equation}\label{Hext}
{\hat H}_{\rm ext} = \frac{{\hat p}_z^2}{2M}+\frac{M \omega_{\rm osc}^2 {\hat z}^2}{2}
\end{equation}
is the motional contribution of a harmonically confined ion with mass $M$ in 1D approximation. Operator ${\hat V}_1$ describes transitions induced by the cooling laser

\begin{equation}\label{E1}
{\bf E}_1 = \frac{{\bf E}_{01}}{2} \exp(ik_1z-i\omega_1 t) +c.c.,
\end{equation}
being in resonance with the $\ket{0}\to\ket{1}$ transition and ${\hat V}_2$ describes the action of the quenching field

\begin{equation}\label{E2}
{\bf E}_2 = \frac{{\bf E}_{02}}{2} \exp(ik_2z-i\omega_2 t) +c.c.,
\end{equation}
resonant with the $\ket{1}\to\ket{2}$ transition, as depicted in fig.~\ref{fig:quantummodel}(a). In the rotating wave basis the Hamiltonian of the internal ion states is given by
\begin{equation}
{\hat H}_{\rm int} = -\delta_2 {\hat P}_2  -\delta_1 {\hat P}_1,
\end{equation}
where ${\hat P}_1 $ and ${\hat P}_2$ are projection operators to the states
$\ket{1}$ and $\ket{2}$. $\delta_1 = \omega_1-\omega_{10}$ and
$\delta_2 = \omega_2-\omega_{21}$ are the detunings of the corresponding light fields (\ref{E1}) and (\ref{E2}), where $\omega_{10}$ and $\omega_{21}$ are the resonance frequencies of the unperturbed $\ket{0}\to\ket{1}$ and $\ket{1}\to\ket{2}$ transitions, respectively.

In the basis of Fock states the interaction operators ${\hat V}_1$ and
${\hat V}_2$ contain components that determine the amplitudes of transitions
between the states with different vibrational numbers. The Rabi frequency induced by the field ($i = 1$ for the cooling light, and $i = 2$ for the quenching light) coupling the states with different vibrational numbers $m$ ($n$) in the ground (excited) states is determined by the expression (see~\cite{Wineland1979,Leibfried2003} for details):

\begin{equation}\label{Rnm}
\Omega_{nm}^{(i)} = \Omega^{(i)} C_{nm}(\eta_i),
\end{equation}
with
\begin{equation}\label{eq:strength_coef}
C_{nm}(\eta_i)=L^{|n-m|}_{n_<}\left(\eta_i^2\right)\sqrt{\frac{n_<!}{n_>!}}\left(i\eta_i\right)^{|n-m|}
\exp\left(-\frac{\eta_i^2}{2}\right),
\end{equation}
where $n_< = \min\{n,m\}$, $n_> = \max\{n,m\}$, and $L^{|n-m|}_{n_<}\left(x \right)$ is the generalized
Laguerre polynomial. A small Lamb-Dicke parameter  $\eta_i = \sqrt{\hbar k^2_i/(2M\omega_{\rm osc})} \ll 1$, i.e. tight harmonic confinement, significantly reduces the probability of induced and spontaneous transitions between energy states with different vibrational numbers $m \neq n$, where the $k_i$ are the wave vectors of the corresponding light fields. The full quantum treatment of the three-level system with the dynamical evolution of the density matrix elements is discussed in~\ref{sec:densitymatrix}.

As shown in~\cite{Marzoli1994}, the action of the quenching field ${\bf E}_2$ onto a three-level system, leading to a fast decay to the ground state, allows for the reduction to an effective two-level system for the $\ket{0}$ and $\ket{1}$ states (see figure~\ref{fig:quantummodel}(a)) with effective decay rate

\begin{equation}\label{geff}
\gamma_{\rm eff} =\gamma_1+\gamma_3 \frac{\rho^{22}}{\rho^{11}}\simeq
\gamma_1\left(1+\frac{\gamma_3}{\gamma_1}S_2 \right),
\end{equation}
where $S_2 = \Omega_2^2/[(\gamma_1+\gamma_2+\gamma_3)^2+4\delta_2^2]$ is the quenching field saturation parameter and the $\gamma_i$ ($i=1..3$) are the decay rates of the involved transitions. A quantitative analysis of the cooling performance is derived from the time evolution of the mean occupation number $\bar{n}$ of the Fock states, which according to~\cite{Javanainen1981}, can be interpolated by an exponential decay
\begin{equation}\label{exp}
\bar{n}(t) = \left(\bar{n}_{\rm ini} - \bar{n}_{\infty} \right)e^{-a\,t} +
\bar{n}_{\infty},
\end{equation}
where $\bar{n}_\infty$ is the mean occupation number in steady-state. Based on the dynamic equations~(\ref{eqn}), we derive the time evolution of $\bar{n}$ for various Rabi frequencies of the cooling light, assuming an effective decay rate $\gamma_{\rm eff}=2\pi\times\unit[50]{kHz}$. The initial conditions correspond to a thermal distribution of Fock states in the electronic ground state with $\bar{n}_{\rm ini}=20$ for an ion secular frequency of $\omega_{\rm osc}/2\pi =\unit[600]{kHz}$. As can be seen in figure~\ref{fig:quantummodel}(b), we expect the existence of an optimal value of cooling laser intensity corresponding to a maximised cooling rate, which is not predicted by~\cite{Javanainen1981}. In the following, we thus investigate the optimal cooling parameters for fastest cooling into the motional ground state in the resolved sideband, intermediate and semi-classical regime.

\section{The ``${\hat \tau}$-matrix method'' for fast simulation of cooling dynamics}
The exponential interpolation in equation~(\ref{exp}) requires to numerically solve the dynamic equations (\ref{eqn}), which, taking into account a large number $n$ of Fock states, requires significant computational resources to solve for a $2n\times 2n$ density matrix for a two-level atom. An alternative approach to derive the cooling time is given by the ``${\hat \tau}$-matrix method'', recently introduced in \cite{Ilenkov2016} for cooling of neutral atoms. The ${\hat \tau}$-matrix is given by the time integral of the difference of the atomic density matrix ${\hat \rho}(t)$ and its steady state solution ${\hat \rho}_{\rm st} ={\hat \rho}(t)|_{t = \infty}$

\begin{equation}\label{taumatrix}
{\hat \tau} = \int_0^{\infty}\left({\hat \rho}(t)-{\hat \rho}_{\rm st}\right)\,dt
\,.
\end{equation}
The basic equation for the $\hat \tau$-matrix is given by
\begin{equation}\label{tau_eq}
-\frac{i}{\hbar}\left[{\hat H},{\hat \tau}\right] +{\hat \Gamma}\left\{  {\hat
\tau} \right\} ={\hat \rho}_{\rm st}-{\hat \rho}_{\rm ini},
\end{equation}
where ${\hat \rho}_{\rm ini}={\hat \rho}(t)|_{t=0}$ is the density matrix at initial time. As the density matrix $\hat{\rho}$ contains all information on external and internal states of the quantum system, the  ${\hat \tau}$-matrix
contains all information on temporal characteristics of the system. As an example for an observable $A$ characterized by the quantum operator $\hat{A}$, the characteristic evolution time can be extracted from the  ${\hat \tau}$-matrix by the following expression \cite{Ilenkov2016}:
\begin{equation}
\tau_{A} = \frac{\mbox{Tr}\left\{{\hat A} \, {\hat
\tau}\right\}}{\left(\mbox{Tr}\left\{{\hat A} \,{\hat
\rho}_{\rm ini}\right\}-\mbox{Tr}\left\{{\hat A} \,{\hat
\rho}_{\rm st}\right\}\right)} \, .
\end{equation}
The cooling time is determined by the evolution rate of the external degrees of freedom determined by operator ${\hat H}_{\rm ext}$ in equation (\ref{Hext}), and thus can be defined as
\begin{equation}\label{tau}
\tau_{\rm eff} = \frac{\mbox{Tr}\left\{{\hat H}_{\rm ext} \, {\hat
\tau}\right\}}{\left(\mbox{Tr}\left\{{\hat H}_{\rm ext} \,{\hat
\rho}_{\rm ini}\right\}-\mbox{Tr}\left\{{\hat H}_{\rm ext} \,{\hat
\rho}_{\rm st}\right\}\right)} \, .
\end{equation}
In the case of a trapped ion this expression can be reduced to
\begin{equation}
\tau_{\rm eff} = \frac{1}{\left(\overline{n}_{\rm ini} - \overline{n}_{\infty}
\right)}\int_0^{\infty}\left(\overline{n}(t)-\overline{n}_{\infty}\right)\,dt
\end{equation}
which exactly corresponds to the $1/e$ value of an exponential decay of $\bar{n}(t)$, as given in equation (\ref{exp}) with characteristic time $\tau=1/a$.
As an example, for the results depicted in figure~\ref{fig:quantummodel}(b) the ``$\hat\tau$-matrix
method'' gives the following values $\tau_{\rm eff} \simeq (115,\,31,\, 55)\,\gamma_{\rm eff}^{-1} $, with Rabi frequencies $\Omega/2\pi = \unit[(50, 200, 500)]{kHz}$, that are in good agreement with the results obtained through a fit of the direct numerical solution $\bar{n}(t)$ by the exponential function (\ref{exp}), $\tau \simeq (115,\,29,\, 50)\,\gamma_{\rm eff}^{-1}$ for corresponding Rabi frequencies. Differences between $\tau$ and $\tau_{\rm eff}$ should
become noticeable if the evolution of $\bar{n}(t)$ is not governed by an exponential decay. 

The ``$\hat{\tau}$-matrix method'' allows to reduce the analysis based on solving the dynamical equations for the density matrix to a more simple task, i.e. the solution of the linear equation (\ref{tau_eq}). As an example, taking into account $n=30$ Fock states, computation of the effective cooling time with the ``$\hat{\tau}$-matrix method'' takes approx. $\unit[45]{s}$ in our case, while the direct solution of eq.~(\ref{dm}) takes $\unit[430]{s}$. This allows us to perform a detailed analysis of the characteristic cooling time of a trapped ion with taking a sufficiently large number of vibrational states ($n_{\rm max} \simeq 120$) into account.

\section{Dynamics of a trapped ion in the sideband cooling regime}
We now use the ``$\hat{\tau}$-matrix method'' for a general study of the cooling dynamics in the resolved sideband cooling regime (where $\gamma_{\rm eff} \ll
\omega_{\rm osc}$) which, in principle, can be applied to any atomic species featuring a level structure as shown in fig.~\ref{fig:quantummodel}(a).

\begin{figure}[t]
    \centering
    \includegraphics[width=\textwidth]{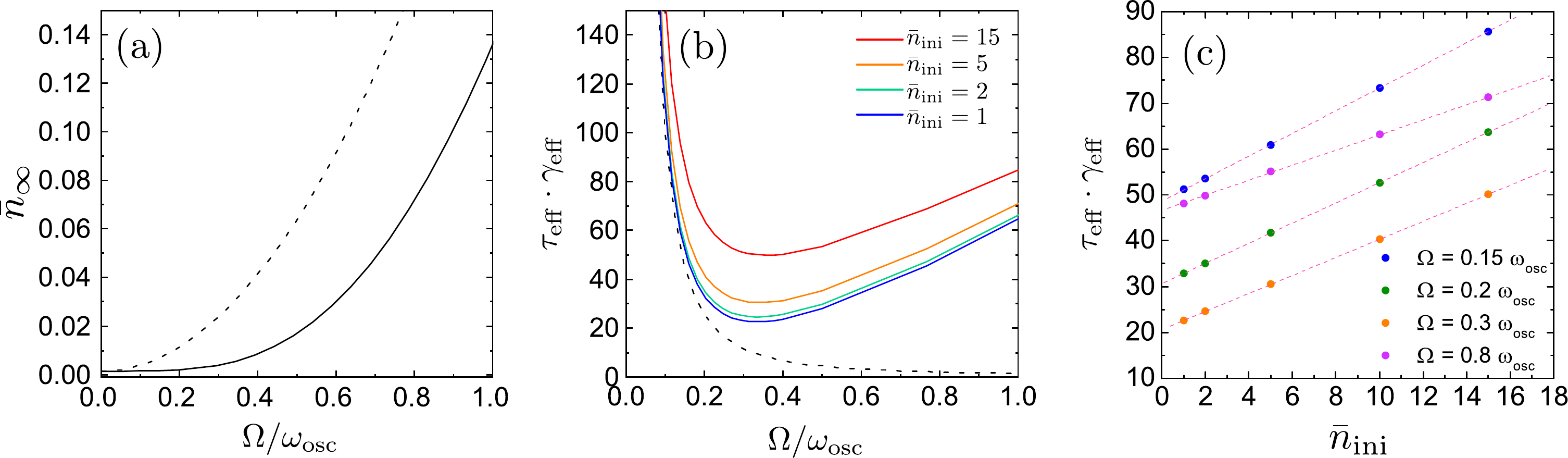}
    \caption{(a) Steady-state mean occupation number and (b) cooling time $\tau_{\rm eff}$ as function of cooling laser Rabi frequency $\Omega$. A minimum cooling time $\tau_{\rm min}$ is observed at optimal Rabi frequency $\Omega_{\tau}$. The dashed lines in (a) and (b) are derived from simplified balance-rate equations given in~\cite{Javanainen1981,Javanainen1984}.  (c) Cooling time $\tau_{\rm eff}$ as function of initial $\bar{n}_{\rm ini}$ for various Rabi frequencies. Quench cooling parameters in (a)-(c) are $\gamma_{\rm eff}/\omega_{\rm osc} = 0.1$, $\eta=0.1$ and detuning $\delta = -\omega_{\rm osc}$. }
    \label{fig:sidebandcooling}
\end{figure}

\begin{figure}[t]
    \centering
    \includegraphics[width=0.8\textwidth]{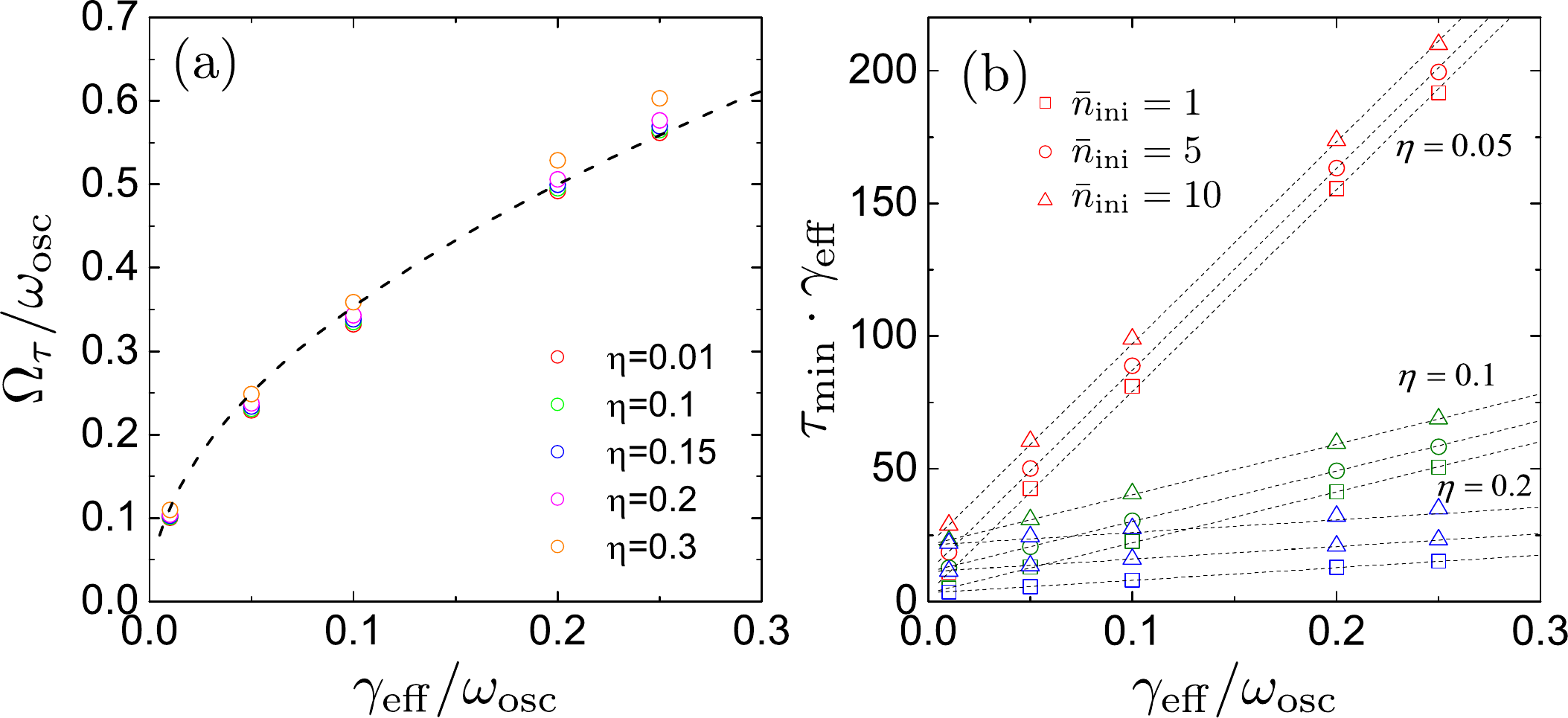}
    \caption{(a) Optimal Rabi frequency $\Omega_{\tau}$ for reaching $\tau_{\rm min}$ as function of $\gamma_{\rm eff}$ for various Lamb-Dicke parameters $\eta$. The black dashed line has been calculated with the analytical expression given by Equation~(\ref{sidebandRabi}). (b) Minimum cooling time at optimal Rabi frequency $\Omega_{\tau}$ as function of effective decay rate for various $\bar{n}_{\rm ini}$ (indicated by different symbols) and Lamb-Dicke parameters $\eta$ (indicated by different colours). The detuning of the cooling laser corresponds to the first red sideband resonance $\delta = -\omega_{\rm osc}$. The black dashed lines have been calculated with the expression for $\tau_{\rm min}$ given in Equation~(\ref{sidebandTau}) and plotted in units of $\gamma_{\rm eff}^{-1}$.  }
    \label{fig:optomega}
\end{figure}

In figures~\ref{fig:sidebandcooling}(a) and (b) we plot the mean occupation number $\bar{n}_\infty$ and the cooling time $\tau_{\rm eff}$, respectively, as a function of Rabi frequency $\Omega$ of the cooling light. We compare our results to calculations derived from simplified balance-rate equations in the low-intensity $S_1\ll1$ ($S_1$ is the saturation parameter of the cooling transition) and strong Lamb-Dicke limit $\eta \ll 1$ (dashed lines), without taking into account the coherence of the Fock states~\cite{Javanainen1981,Javanainen1984}. Obtaining similar results for the low-intensity limit, our calculations indicate a global minimum in cooling time, $\tau_{\rm min}$, at Rabi frequencies between $\Omega_{\tau}=0.3\,\omega_{\rm osc}$ and $\Omega_{\tau}=0.4\,\omega_{\rm osc}$. This point defines the optimal parameters for both fast and simultaneously deep laser cooling, as the mean occupation number has not significantly changed from its minimum value in the low-intensity limit. Furthermore, we observe a linear dependence of cooling time $\tau_{\rm eff}$ on the initial mean occupation number $\bar{n}_{\rm ini}$, as shown in Figure~\ref{fig:sidebandcooling}(c). As a next step, we study the dependence of the optimal parameters on the effective quench rate $\gamma_{\rm eff}$. As shown in figure~\ref{fig:optomega}(a), we observe that $\Omega_{\tau}$ strongly depends on $\gamma_{\rm eff}$ in the sideband cooling regime, but does not much depend on the Lamb-Dicke parameter $\eta$. The available amount of simulation data for cooling in the sideband regime with parameters $\gamma_{\rm eff}/\omega_{\rm osc}<0.5$, $\eta < 0.3$ and $\bar{n}_{\rm ini}< 20$ allows us to deduce analytical expressions from fit results for the optimal Rabi frequency $\Omega_{\tau}$ 
\begin{equation}\label{sidebandRabi}
\Omega_{\tau} \simeq \sqrt{1.24\cdot\gamma_{\rm eff}\cdot\omega_{\rm osc}} 
\end{equation}
and the minimum cooling time $\tau_{\rm min}$ at $\Omega_{\tau}$ and detuning $\delta = -\omega_{\rm osc}$:

\begin{equation}\label{sidebandTau}
\tau_{\rm min} \simeq \frac{1.2+2\,\bar{n}_{\rm ini} }{\gamma_{\rm eff}} +
\frac{1.9}{\eta^2\omega_{\rm osc}}.
\end{equation}
Note that these are general expressions valid for any two-level ion confined in a harmonic trap with a decay rate $\gamma_{\rm eff}$ and Lamb-Dicke parameter $\eta$. We use equations~(\ref{sidebandRabi}) and~(\ref{sidebandTau}) to plot the dashed black lines in figure~\ref{fig:optomega} and observe good agreement with the results for $\Omega_{\tau}$ and $\tau_{\rm min}$ obtained with our direct simulation. In particular, according to equation (\ref{sidebandRabi}), the optimal Rabi frequency providing fast cooling for parameters used in figure \ref{fig:quantummodel}(b) results to $\Omega_{\tau}/2\pi \simeq 193$ kHz, which is in excellent agreement with the result of the direct simulation of cooling dynamics shown there.

\section{Analysis of various quench cooling regimes in $^{172}$Yb$^+$}
\begin{figure}[h]
    \centering
    \includegraphics[width=0.4\textwidth]{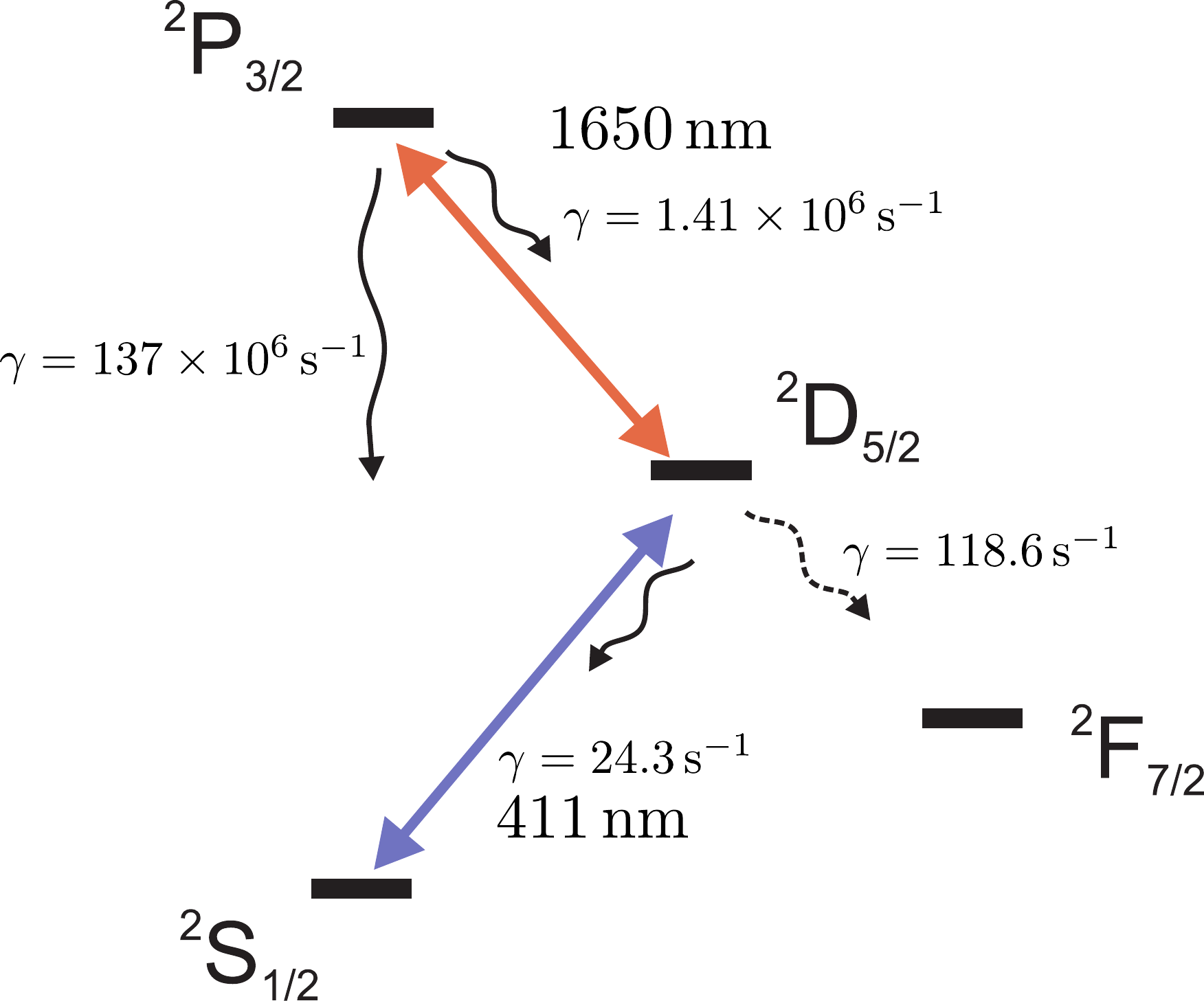}
    \caption{Relevant atomic levels for quench cooling in $^{172}$Yb$^+$ including their decay rates. Sideband cooling is carried out on the electronic $^2$S$_{1/2}\to\,^2$D$_{5/2}$ quadrupole transition near $\unit[411]{nm}$. The $^2$D$_{5/2}$ state is coupled to the short-lived $^2$P$_{3/2}$ state by means of laser light near $\unit[1650]{nm}$.  }
    \label{fig:levelscheme}
\end{figure}

Turning from a general treatment to a specific case, we now apply our ``$\hat{\tau}$-matrix method'' to a $^{172}$Yb$^+$ ion confined in a rf Paul trap. Figure~\ref{fig:levelscheme} shows the relevant atomic states involved in the quench cooling process. The $\ket{0}\to\ket{1}$ cooling transition is given by the $^2$S$_{1/2}\to\,^2$D$_{5/2}$ quadrupole transition near $\unit[411]{nm}$, where the $^2$D$_{5/2}\to\,^2$P$_{3/2}$ dipole transition near $\unit[1650]{nm}$ is used as $\ket{1}\to\ket{2}$ quenching transition. In the following, we compare the rates and limits of quench cooling in the resolved sideband regime $\gamma_{\rm eff} \ll \omega_{\rm osc}$, the semi-classical regime $\gamma_{\rm eff}>\omega_{\rm osc}$ and in the intermediate cooling regime $\gamma_{\rm eff} \simeq \omega_{\rm osc}$.   

\begin{figure}[h]
    \centering
    \includegraphics[width=\textwidth]{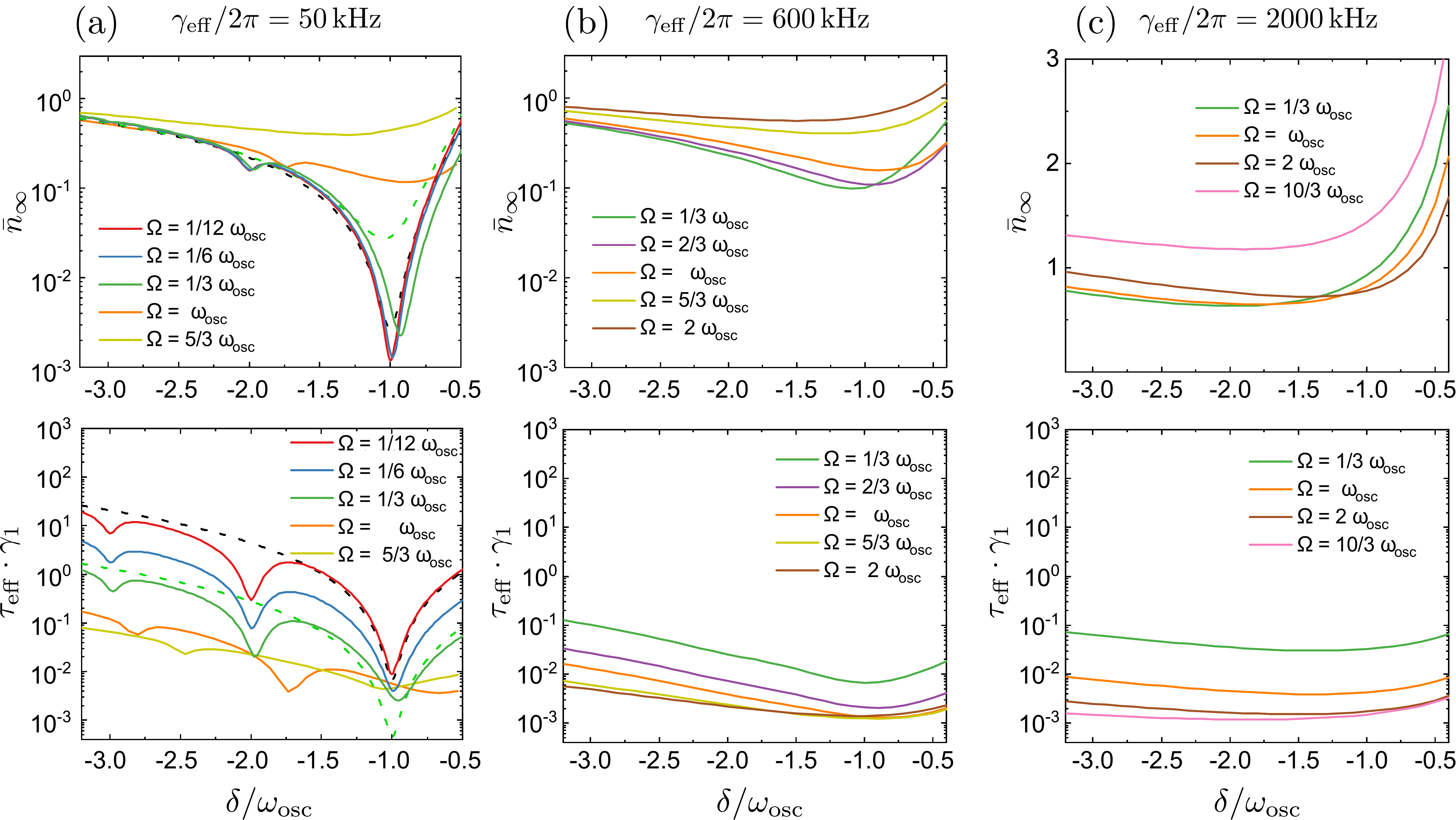}
    \caption{Comparison of quench cooling in (a) the sideband cooling regime with $\gamma_{\rm eff}/2\pi=\unit[50]{kHz}$, (b) intermediate cooling regime with $\gamma_{\rm eff}/2\pi=\unit[600]{kHz}$ and (c) semi-classical cooling regime with $\gamma_{\rm eff}/2\pi=\unit[2000]{kHz}$. For each of the regimes, both $\bar{n}_\infty$ and the cooling time $\tau_{\rm eff}$ (in units of the decay rate $\gamma^{-1}_1$ of the $^2$S$_{1/2}\to\,^2$D$_{5/2}$ transition) are plotted for various Rabi frequencies of the cooling light field ${\bf E}_1$. All cooling laser parameters ($\Omega$ and $\delta$) are given in units of ion secular frequency $\omega_{\rm osc}=2\pi\times\unit[600]{kHz}$. Each cooling simulation started with an initial Boltzmann distribution with $\bar{n}_{\rm ini}=20$. The dashed black and green curves in (a) correspond to $\bar{n}_\infty$ and cooling time at $\Omega/\omega_{\rm osc}=1/12$ and $1/3$ obtained from analytical expressions given in~\cite{Javanainen1981}.}
    \label{fig:regimes}
\end{figure}

For each of these regimes, Figure~\ref{fig:regimes} shows the results for the mean occupation number $\bar{n}_\infty$ and the cooling time for different Rabi frequencies induced by the cooling field ${\bf E}_1$. In the resolved sideband regime, we observe a minimum of ${\bar n} \simeq 0.0012$ for a detuning $\delta=-\omega_{\rm osc}$ at low intensities ($\Omega = 1/12\,\omega_{\rm osc}$), which is close to the limit $\bar{n}_{\rm min} \simeq 7/48 \,(\gamma_{\rm eff}/\omega_{\rm osc})^2 \simeq 0.001$ at low intensity as was obtained similarly to~\cite{Javanainen1981,Javanainen1984}, or $\bar{n}_{\rm min} \simeq
5/16\, (\gamma_{\rm eff}/\omega_{\rm osc})^2\simeq 0.002$ in \cite{Wineland1979}. Only the linear dependence on light field intensity was taken into account for $\bar{n}$ in the
simplified balance-rate equations~\cite{Javanainen1981,Javanainen1984}, which gives slightly underestimated results for the mean occupation number compared to our direct simulation obtained by the ``${\hat \tau}$-matrix method''. For visibility, we used a logarithmic scale in Figure~\ref{fig:regimes}(a) to pronounce the differences at low intensities (black dashed and solid curves). The results for the cooling time coincide with \cite{Javanainen1981,Javanainen1984} in the vicinity of $\delta = -\omega_{\rm osc}$ at low laser intensity. However, with increasing intensity the difference between the simple model and our direct simulation becomes more pronounced (green dashed and solid lines for $\Omega = 1/3\, \omega_{\rm osc}$). We also observe that the optimal detuning for  minimum $\bar{n}_{\infty}$ and cooling time shifts to smaller absolute values. Additionally, at low intensity, local minima are also visible near $\delta/\omega_{\rm osc} = -2,-3,\ldots$. These effects are not predicted by the simplified model \cite{Javanainen1981,Javanainen1984}. 


Note the different dependencies of temperature and cooling time on cooling laser intensity: in the weak-field limit, the ion temperature (or the mean occupation number $\bar{n}$) does not significantly depend on intensity, but cooling time decreases proportional to it. For high-intensity laser fields the temperature significantly grows, but does not result in essential further decreasing of cooling time. Such a dependence of cooling time and temperature on field intensity allows to define an optimal cooling light intensity for effective, i.e. fast and simultaneously deep quench cooling. A similar behaviour can be observed in the intermediate (figure~\ref{fig:regimes}(b)) and semi-classical cooling regime (figure~\ref{fig:regimes}(c)). 

\begin{figure}[h]
    \centering
    \includegraphics[width=\textwidth]{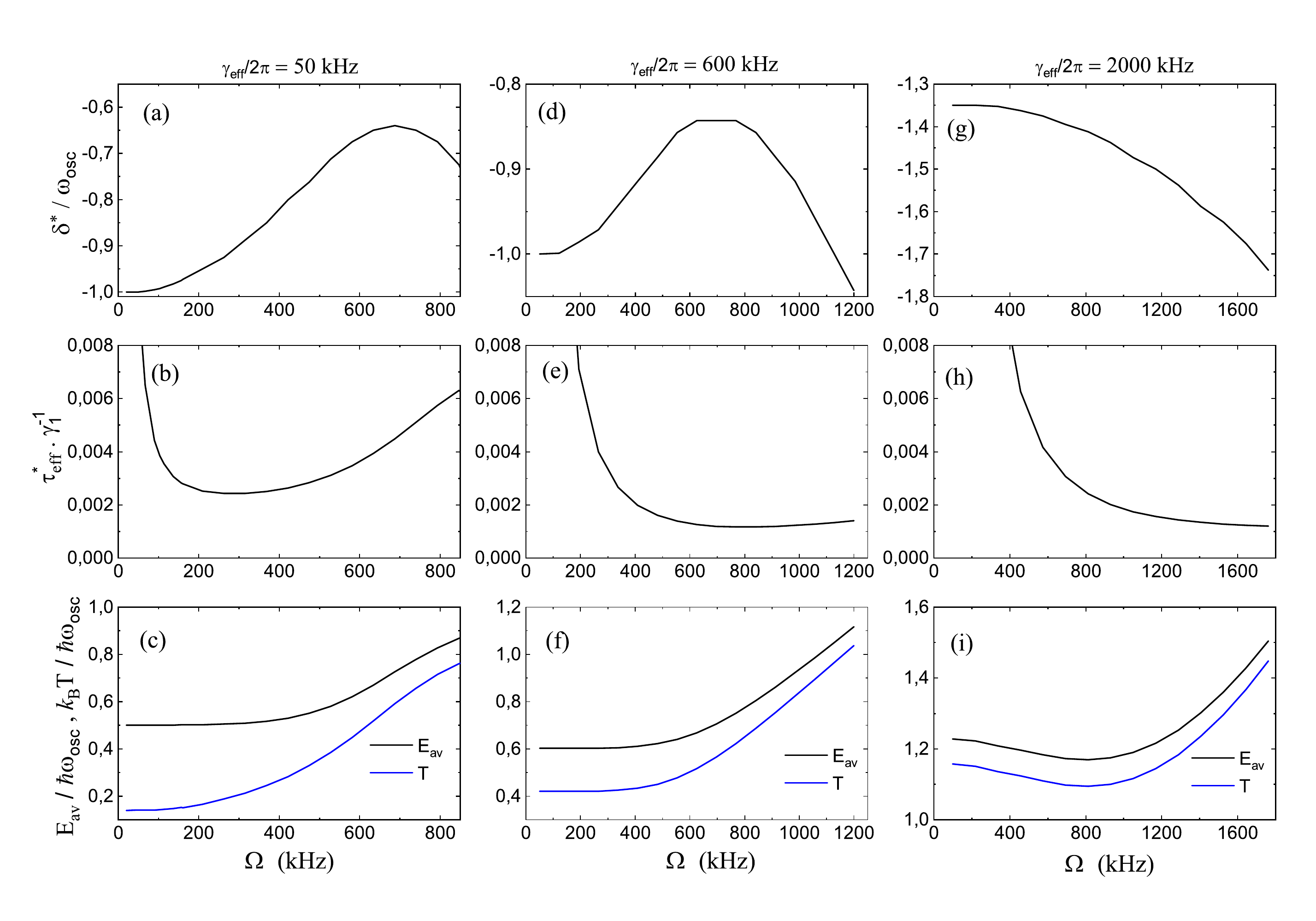}
    \caption{Optimal parameters for fastest quench cooling in various regimes: resolved sideband regime (a, b, c), intermediate regime (d, e, f), and semi-classical cooling regime (g, h, i). In each of the cases, the optimal detuning $\delta^*$ will lead to an optimal cooling time $\tau^*$ (in units of the decay rate $\gamma^{-1}_1$ of the $^2$S$_{1/2}\to\,^2$D$_{5/2}$ transition). The ion temperature $T$ is obtained by fitting the steady-state population governed by $\hat\rho_{\rm st}$ with an exponential distribution. Using this $T$ in a Boltzmann distribution results to the average cooling energy $E_{\rm av}$. In each of the cooling regimes shown here, the ion secular frequency was $\omega_{\rm osc}/2\pi=\unit[600]{kHz}$ and the simulations started with an initial thermal distribution with $\bar{n}_{\rm ini} = 20$.}
    \label{fig:optimalparameters}
\end{figure}

To select the optimal cooling parameters we use the following algorithm: for various intensities of the cooling light field, we determine the optimal detuning $\delta^{*}$ providing the maximum cooling rate. At these values of $\delta^{*}$, we then analyse the steady-state temperature. The results are shown in figure~\ref{fig:optimalparameters}. The optimal detuning $\delta^{*}$ shifts away from the sideband resonance $\delta=-\omega_{\rm osc}$ with growing intensity (figures~\ref{fig:optimalparameters}(a, d, g)). In each of the studied cooling regimes, a minimum for the cooling time can be observed in a certain range of cooling light intensity (figures~\ref{fig:optimalparameters}(b, e, h)), which allows to select the optimal intensity for fast cooling. In figures~\ref{fig:optimalparameters}(c, f, i) we plot the steady-state temperature $T$ and average atomic energy $E_{\rm av}$ as a result from cooling at optimal detuning $\delta^{*}$. The steady-state temperature, which we plot in units of $\hbar\omega_{\rm osc}/k_{\rm B}$, is determined by fitting an exponential distribution to the steady-state populations governed by $\hat\rho_{\rm st}$. The average atomic energy results as the mean value $E_{\rm av}=\sum_n nP_n$, where $P_n$ is a Boltzmann distribution with steady-state temperature $T$ over the atomic states with secular frequency $\omega_{\rm osc}/2\pi=\unit[600]{kHz}$. Note that $E_{\rm av}$ is plotted in units of $\hbar\omega_{\rm osc}$ with offset of $0.5$ to account for the quantum mechanical ground state energy.    

The absolute minimum cooling temperature is reached in the resolved sideband cooling regime $\gamma_{\rm eff} \ll \omega_{\rm osc}$. The corresponding minimum cooling time $\tau^{*} \simeq 2.4\times 10^{-3} \,\gamma^{-1}_1\simeq \unit[100]{\mu s}$ at optimal detuning $\delta^{*}\simeq\unit[-0.89]{\omega_{\rm osc}}$ is reached for a Rabi frequency $\Omega^{*} \simeq \omega_{\rm osc}/2$ of the cooling light. Here, the steady-state temperature is only slightly larger than the minimum value obtained in the low-intensity limit. Both in the intermediate and semi-classical cooling regime, the minimum cooling time is found for a Rabi frequency of the cooling light
\begin{equation}\label{opt}
\Omega^{*} \simeq \sqrt{ \gamma^2_{\rm eff}/2+2\,\omega^2_{\rm osc} } \,,
\end{equation}
where the ion temperature increases proportionally with increasing Rabi frequency $\Omega^{*}$. Note that the analytical approximations for the optimum Rabi frequency are again very general and can be applied to quench cooling in atomic species with similar level schemes. 

To draw a first conclusion on our findings, the intermediate regime is very promising for fast laser cooling to $\bar{n}<1$, as the observed minimum cooling time $\tau^{*} \simeq 1.2\times 10^{-3} \, \gamma^{-1}_1 \simeq\unit[50]{\mu s}$ is very similar to the semi-classical cooling regime, but at the expense of slightly increased temperature.  

We benchmark our simulation with the ``$\hat\tau$-matrix method'' against data acquired with a single $^{172}$Yb$^+$ ion confined in a high-precision rf Paul trap~\cite{Keller2019}. The rf drive frequency is $\Omega_{\rm rf}=2\pi\times\unit[24.4]{MHz}$ leading to a typical secular frequency of the strong radial mode of $\omega_x=2\pi\times\unit[565(5)]{kHz}$. In a first stage, the ion is Doppler-cooled to $T_{\rm D}\sim\unit[0.5]{mK}$ on the $^2$S$_{1/2}\to\,^2$P$_{1/2}$ transition near $\unit[370]{nm}$. The resulting thermal distribution of Fock states features a mean occupation number $\bar{n}=18(1)$, measured with Rabi flops recorded on the $^2$S$_{1/2}\to\,^2$D$_{5/2}$ transition near $\unit[411]{nm}$. 

\begin{figure}[h]
    \centering
    \includegraphics[width=0.6\textwidth]{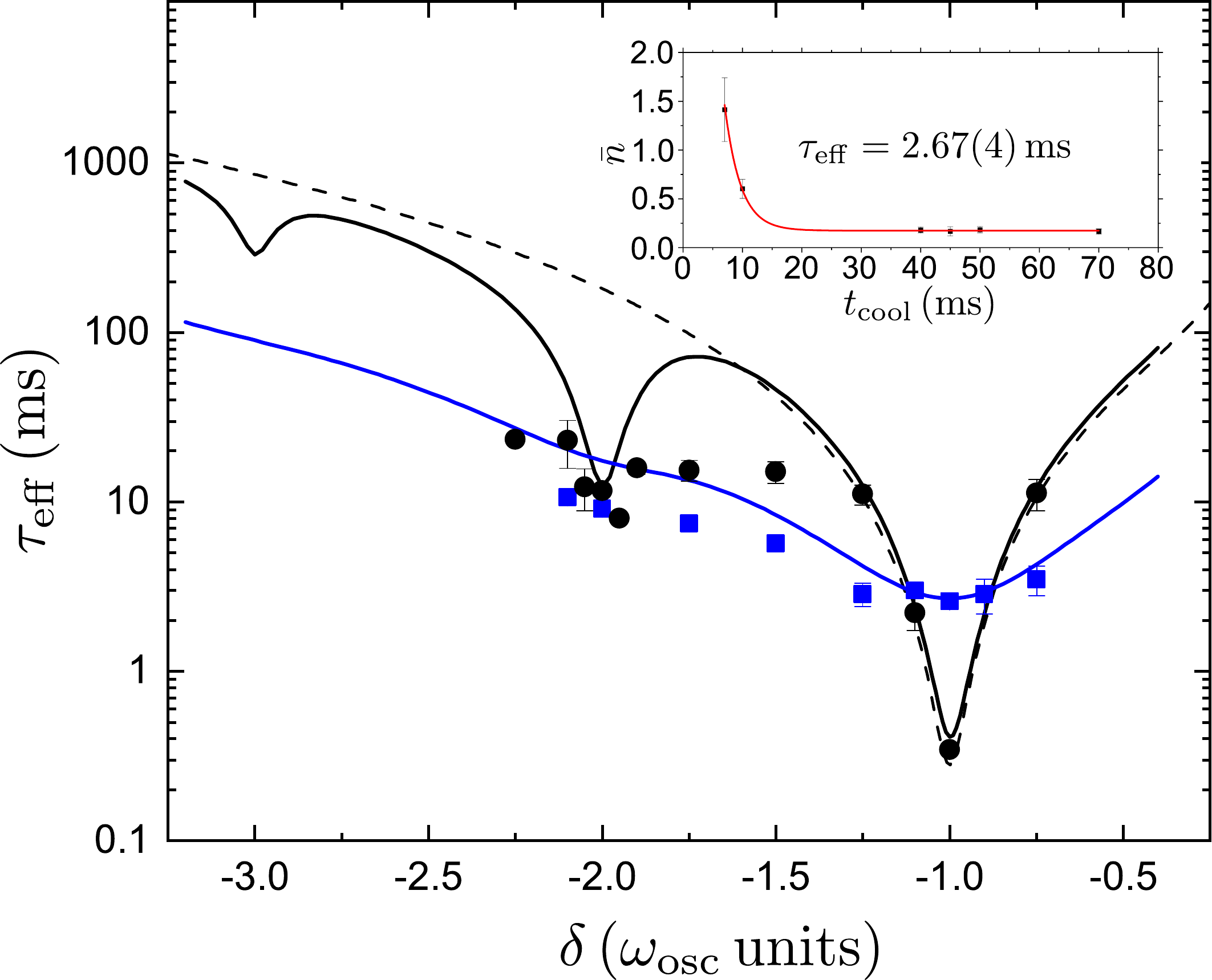}
    \caption{Effective cooling time $\tau_{\rm eff}$ for quench cooling in the resolved sideband regime with $\gamma_{\rm eff}/2\pi=\unit[48(2)]{kHz}$ (black) and close to the intermediate regime with $\gamma_{\rm eff}/2\pi=\unit[404(21)]{kHz}$ (blue) as a function of cooling laser detuning in units of ion secular frequency $\omega_{\rm osc}=2\pi\times\unit[565(5)]{kHz}$. The dashed line is derived from analytical expressions in~\cite{Javanainen1981} and the solid curves correspond to our direct simulation with $\Omega=2\pi\times\unit[50]{kHz}$. The filled squares and circles are experimental data. Each experiment was carried out with a laser intensity corresponding to $\Omega=2\pi\times\unit[50(2)]{kHz}$ and an initial thermal distribution with $\bar{n}\simeq 18(1)$. The inset exemplarily shows how the effective cooling time for $\gamma_{\rm eff}/2\pi=\unit[404(21)]{kHz}$ and $\delta=-\omega_{\rm osc}$ was determined.    }
    \label{fig:detuning}
\end{figure}

Following the Doppler cooling stage, the ion is initialized in the $m_J=-1/2$ ground state. The cooling laser near $\unit[411]{nm}$ is set to a Rabi frequency $\Omega/2\pi=\unit[50(2)]{kHz}$ and detuning $\delta$ where we derive the ion temperature as a function of quench cooling time $t_{\rm cool}$ via the amplitude ratio $R=I_{\rm BSB}/I_{\rm RSB}$ of blue and red sidebands~\cite{Diedrich1989}. The mean occupation number is then given by
\begin{equation}\label{sideband}
\bar{n}_{\rm SB}=\frac{1}{R-1}.
\end{equation}
Each individual data point of the sideband scans is repeated 200 times for significant statistics. The effective cooling time $\tau_{\rm eff}$ is derived as a decay parameter from an exponential fit, as shown in the inset of figure~\ref{fig:detuning}. We characterized the cooling rate in a frequency interval ranging from $-0.75$ to $-2.25\,\omega_{\rm osc}$ spanning at least the first two red secular sidebands, i.e. $\delta=-\omega_{\rm osc}$ and $\delta=-2\omega_{\rm osc}$, respectively. The impact of sideband cooling is well pronounced for a moderate quench with $\gamma_{\rm eff}/2\pi=\unit[50(2)]{kHz}$ (black curve) and our measurement is in excellent agreement with the direct simulation in close vicinity of $-\omega_{\rm osc}$ and $-2\,\omega_{\rm osc}$. However, we observe a faster cooling in the region between the first and second red sideband, which we attribute to off-resonant excitation of the sidebands due to laser noise at $\unit[350]{kHz}$ (with a FWHM of approx. $\unit[170]{kHz}$, see also Figure~\ref{fig:noise} in~\ref{sec:lasernoise}), leading to additional cooling. As expected, the sideband signature vanishes for cooling close to the intermediate regime with $\gamma_{\rm eff}/2\pi=\unit[404(21)]{kHz}$ (blue curve). As the cooling laser Rabi frequency is limited by available laser power, we are not able to resolve an even faster cooling rate in this regime, as predicted in figure~\ref{fig:optimalparameters}(e).   


\section{Fock state distribution during quenched sideband cooling}
\label{sec:distribution}
A distribution of Fock states being highly non-thermal after cooling is known to cause a significant underestimate of temperature~\cite{Che2017,Chen2017,Rasmusson2021}. Here, we simulate the distribution of the ion Fock states for various cooling times in our trap using a Monte-Carlo simulation, similar to~\cite{Moelmer1993}, as our previously introduced ``$\hat{\tau}$-matrix method'' only provides the population distribution in steady-state. As we experimentally deduce the ion temperature from the sideband amplitude ratio $R$, we model the red and blue sideband strengths based on the simulated Fock state distribution and derive a corresponding theory value for the temperature according to equation~(\ref{sideband}). For an arbitrary Fock state distribution $P_n$ the sideband amplitude ratio $R$ can be expressed as follows:
    \begin{equation} \label{eq:SB_ratio_general}
        R(t) = \frac{1-\sum_n P_n \cos{\left( C_{n+1, n}(\eta) \Omega_0 t \right)}}{1-\sum_n P_n \cos{\left( C_{n-1, n}(\eta) \Omega_0 t \right)}},
    \end{equation}
where $C_{n+1, n}(\eta)$ and $C_{n-1, n}(\eta)$ are the sideband strength coefficients of the blue and red sideband,  respectively, as given in equation~(\ref{eq:strength_coef}). Note that in general the ratio $R$ is a function of excitation pulse time which is not the case for $P_n$ corresponding to the thermal distribution. For the thermal distribution, the equation~(\ref{eq:SB_ratio_general}) directly implies the well known relation for~$\bar{n}$ shown in expression~(\ref{sideband}).

\begin{figure}[t]
    \centering
    \includegraphics[width=\textwidth]{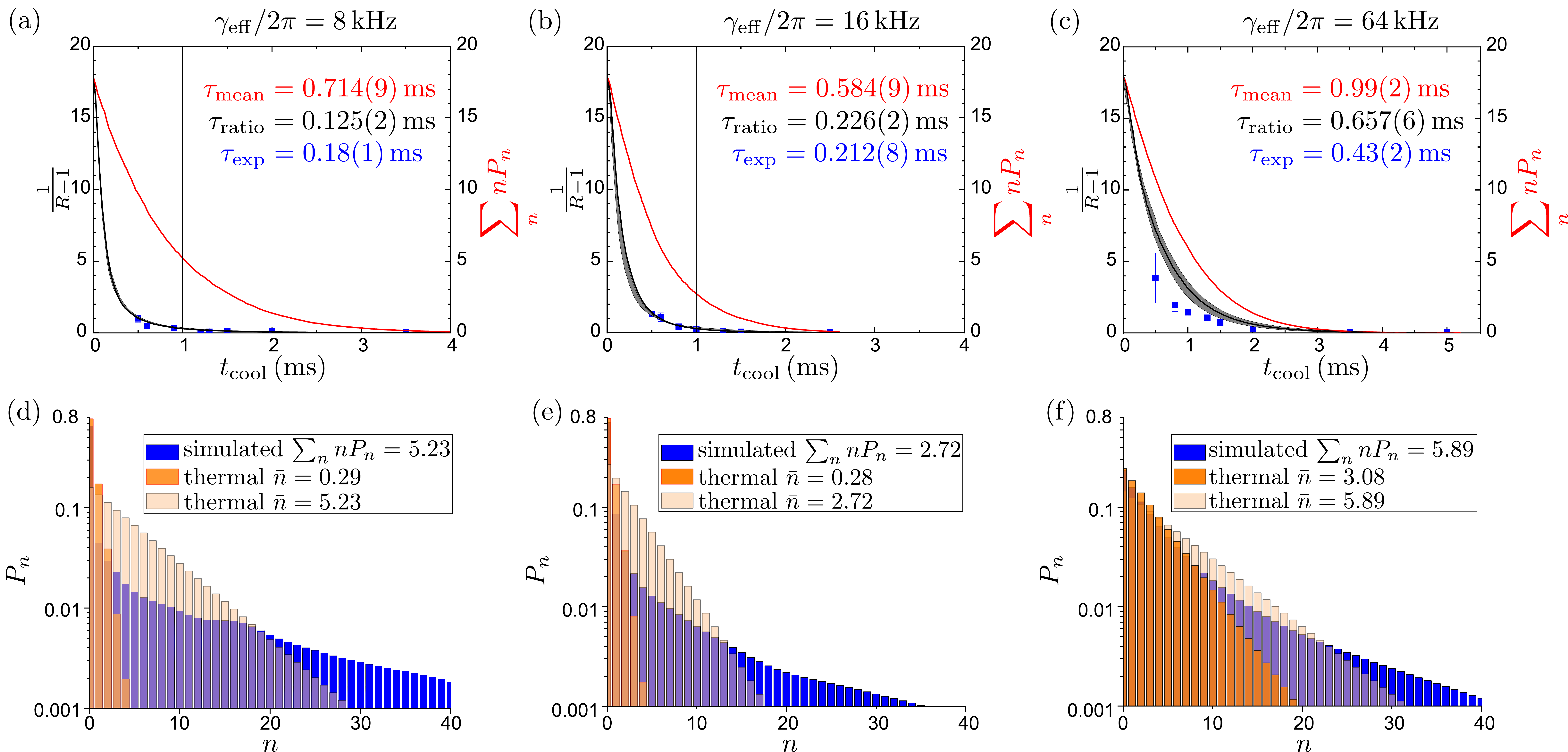}
    \caption{Impact of a non-thermal distribution of atomic Fock states on temperature evaluation. A Monte-Carlo simulation was used to calculate the population distribution for various cooling times, if quench cooling is carried out with $\gamma_{\rm eff}/2\pi=\unit[8,16,64]{kHz}$. For each value of $t_{\rm cool}$ this Fock state distribution is used to calculate the strengths of red and blue sidebands, if driven with a Rabi frequency of $\Omega_{411}=2\pi\times\unit[40]{kHz}$. In (a)-(c) the calculated sideband ratio (black curves) and the mean value of the distribution (red curves) are plotted as a function of cooling time $t_{\rm cool}$. The blue squares correspond to experimentally deduced sideband ratios acquired with (a) $\gamma_{\rm eff}/2\pi=\unit[8.2(2)]{kHz}$, (b) $\gamma_{\rm eff}/2\pi=\unit[16.4(7)]{kHz}$ and (c) $\gamma_{\rm eff}/2\pi=\unit[62(4)]{kHz}$. The shaded region with the black curves account for the uncertainty in our experimental Rabi frequency $\Omega_{411}=2\pi\times\unit[40(2)]{kHz}$. An exponential decay was fitted to each theory curve and experimental data to determine the characteristic cooling time with initial temperature $\bar{n}_{\rm ini}=18$. The fit results are given in (a)-(c). (d)-(f) show - exemplarily at $t_{\rm cool}=\unit[1]{ms}$ - the corresponding simulated Fock state distribution (blue) and thermal distributions with $\bar{n}=1/(R-1)$ (orange) and $\bar{n}=\sum_{n}nP_n$ (beige).  }
    \label{fig:nonthermal}
\end{figure}

Figures~\ref{fig:nonthermal}(a) to (c) show the temperature as a function of cooling time for various effective linewidths $\gamma_{\rm eff}/2\pi$, both plotted for the theoretically obtained $\bar{n}_{\rm SB}$ (black curves) and the mean value of the distribution $\sum_n nP_n$ (red curves). As soon as sideband cooling is initiated, the data obtained from the measurements (blue squares) indeed reveal a non-thermal distribution of Fock states for small effective linewidths (figs.~\ref{fig:nonthermal}(a) and (b)), in good agreement with the calculated temperature $\bar{n}_{\rm SB}$. Note that the $1/e$ cooling times of $\tau_{\rm ratio}=\unit[0.125(2)]{ms}$ and $\tau_{\rm mean}=\unit[0.714(9)]{ms}$ derived for $\bar{n}_{\rm SB}$ and the mean value of the distribution, respectively, significantly differ from each other. In order to pronounce the underestimation of atomic temperature and cooling times, we plot the thermal distributions for $\bar{n}=\bar{n}_{\rm SB}$ (orange) and $\bar{n}=\sum_nn\cdot P_n$ (beige) in figs.~\ref{fig:nonthermal}(d) to (f) together with the previously simulated Fock state distribution (blue). For larger effective linewidths, i.e., $\gamma_{\rm eff}/2\pi=\unit[64]{kHz}$, the distribution of Fock states slowly resembles a thermal distribution, as can be seen in figs.~\ref{fig:nonthermal}(c) and (f). However, we still observe a slightly stronger non-thermal distribution in the experiment. In any case, thermal equilibrium of $\bar{n}\simeq 0.06(2)$ is always reached for sufficiently long cooling times. Our findings suggest that the intermediate regime, where $\gamma_{\rm eff}\simeq\omega_{\rm osc}$, is not only well suited for fast and deep quench cooling to $\bar{n}<1$, moreover it is expected to be rather immune to non-thermal distributions of Fock states during cooling. 

\begin{figure}[h]
    \centering
    \includegraphics[width=\textwidth]{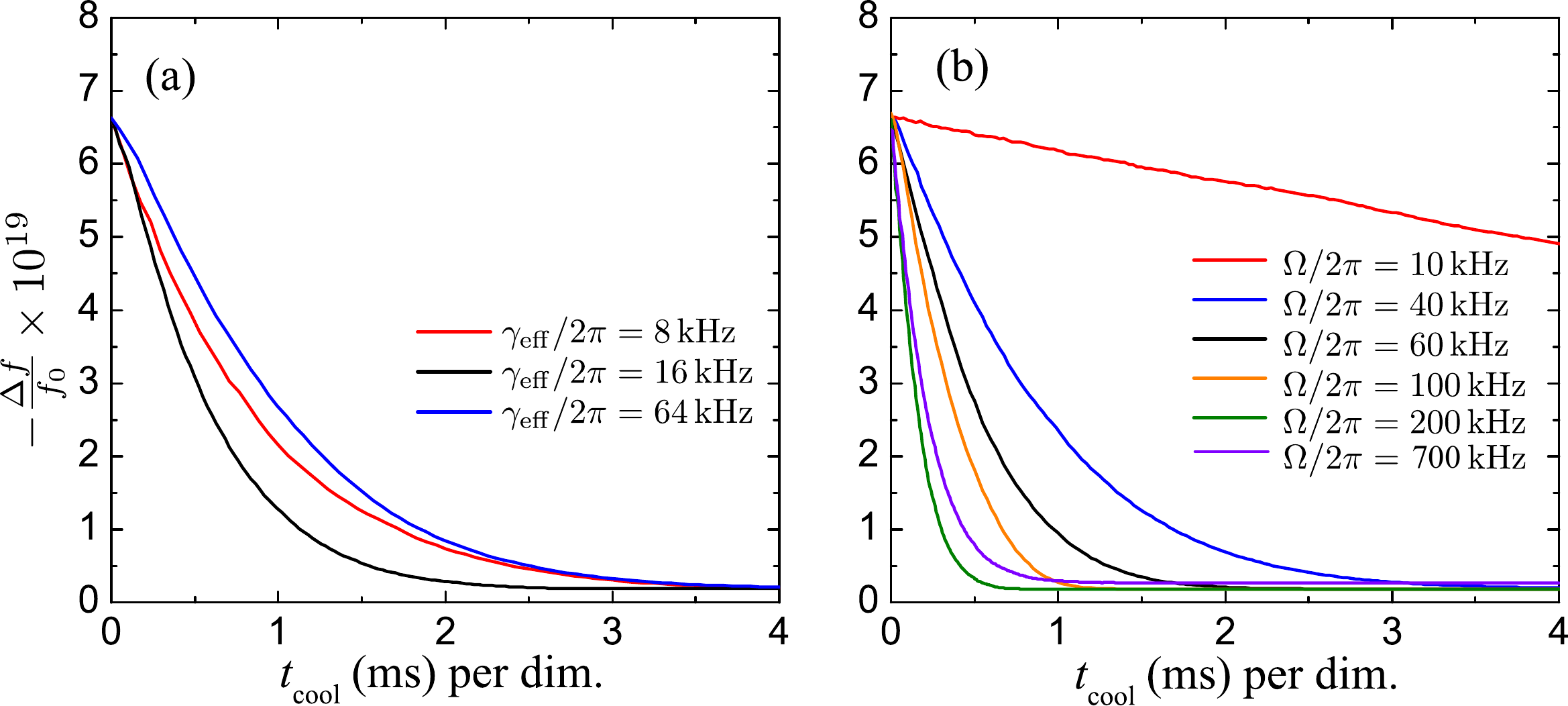}
    \caption{3D relative time dilation shift experienced by a $^{172}$Yb$^+$ ion confined in a linear Paul trap. Here we assume a secular frequency of $\omega_{\rm osc}=2\pi\times\unit[600]{kHz}$ for each motional mode and plot the time dilation as a function of quench cooling time per dimension.  The simulated mean value of the distribution $\sum_{n}nP_n$ amongst the Fock states, as shown in blue in figure~\ref{fig:nonthermal}, was used as a basis for the shift calculation. (a) The Rabi frequency was set to $\Omega/2\pi=\unit[40]{kHz}$ whilst the effective quench rate was varied. (b) For an effective quench rate $\gamma_{\rm eff}=\unit[64]{kHz}$ we observe an optimal Rabi frequency of $\Omega/2\pi=\unit[200]{kHz}$ allowing for fast quench cooling to the point of thermal equilibrium.}
    \label{fig:doppler}
\end{figure}

Being of high importance for optical clocks with trapped ions, we calculate the temporal evolution of thermal time dilation as experienced by a $^{172}$Yb$^+$ ion confined in a linear Paul trap. As a basis for kinetic energy we use the mean $\sum_n nP_n$ of the simulated distribution of Fock states. Figure~\ref{fig:doppler}(a) shows the 3D relative time dilation shift $\Delta f/f_0=-5/2E_{\rm kin}/mc^2$~\cite{herschbach2012}, as a function of cooling time per motional mode with $\omega_{\rm osc}=2\pi\times\unit[600]{kHz}$ for various effective quench rates. In each case, the initial time dilation shift corresponds to the temperature after Doppler cooling. Unless cooling is carried out to thermal equilibrium, we observe a significant dependence of the time dilation shift on the effective quench rate. If cooling was stopped before a thermal equilibrium is reached, as depicted in figures~\ref{fig:nonthermal}(d)-(f), temperature would be falsely determined with the sideband ratio and accordingly lead to a wrong estimation of the time dilation shift. For example, if cooling is carried out for $\unit[1]{ms}$ in the sideband regime with $\gamma_{\rm eff}=\unit[8]{kHz}$ (figure~\ref{fig:nonthermal}d), the sideband ratio would suggest a time dilation shift of $-2.3\times 10^{-20}$ (according to $\bar{n}=0.29$), while the real distribution of Fock states reveals a shift of $-2.2\times 10^{-19}$ (according to $\bar{n}=5.23$). In this regime, the error budget of an atomic clock would be underestimated by one order of magnitude. For reduced discrepancies, it is advised to cool in a regime where $\gamma_{\rm eff}/\omega_{\rm osc}>0.1$, as indicated in figure~\ref{fig:nonthermal}(f). Finally, we would like to point out that the time dilation shift seen by the ion is highly dependent on the cooling laser intensity. In figure~\ref{fig:doppler}(b) we fixed the effective quench rate to $\gamma_{\rm eff}=\unit[64]{kHz}$ and studied the impact of various cooling laser intensities. In accordance with our observation of an optimal Rabi frequency for fastest cooling, as shown in figures~\ref{fig:quantummodel}(b) and~\ref{fig:optimalparameters}(b) , this value also allows to reach the point of thermal equilibrium as fast as possible.

\section{Summary}
To conclude, we studied the mechanism of quenched sideband cooling and presented a versatile method for fast calculation of the characteristic cooling time which does not require to consider the dynamical evolution of the system's density matrix elements. Our ``$\hat{\tau}$-matrix method'' significantly reduces computational efforts and agrees within $90-100\%$ with the full quantum model. 

Based on our powerful simulation tool, we derived universal analytical expressions for the optimum Rabi frequency $\Omega_\tau$ and the resulting minimum cooling time $\tau_{\rm min}$ for cooling in the resolved sideband regime, which can be applied to any atomic species with decay rate $\gamma_{\rm eff}$ confined in a rf Paul trap. Applying the simulation to a specific case, we investigated quench cooling in a $^{172}$Yb$^+$ ion, thereby focussing on three different regimes: (I) resolved sideband cooling with $\gamma_{\rm eff}/2\pi=\unit[50]{kHz}$, (II) intermediate cooling with $\gamma_{\rm eff}/2\pi=\unit[600]{kHz}$ and (III) semi-classical cooling with $\gamma_{\rm eff}/2\pi=\unit[2000]{kHz}$. For each of these regimes we derived the steady-state temperature and the effective cooling time for various cooling laser parameters, such as Rabi frequency and detuning. From this extensive analysis we extracted the optimal parameters to be applied for fast cooling into the motional ground state in each of the aforementioned regimes. We benchmarked our simulation code against data taken with a single $^{172}$Yb$^+$ ion confined with a secular frequency $\omega_{\rm osc}=2\pi\times\unit[565(5)]{kHz}$ and observed an agreement between experiment and theory. 

We presented a detailed analysis of Fock state population distributions during the cooling process. In particular, we compared the time behaviour of temperature derived with the sideband ratio method to the actual mean occupation of Fock states and revealed discrepancies of more than one order of magnitude, if cooling is not carried out to thermal equilibrium. In addition, we investigated the temporal evolution of thermal time dilation in an optical clock with $^{172}$Yb$^+$ during the process of quench cooling. We conclude that quenching in a regime with $\gamma_{\rm eff}/\omega_{\rm osc}>0.1$ is necessary to stay close to a thermal distribution of Fock states.

The results presented in this work pave the way to flexible quench cooling of ion Coulomb crystals with respect to fast laser cooling into the ion's motional ground state. A crystal consisting of $N$ ions requires $3N$ motional modes to be cooled, thus a clever combination of confinement parameters and quench rate is expected to reduce the cooling time compared to cooling of each mode individually. 

\ack We thank Jonas Keller for useful comments on optical spectroscopy in the presence of laser noise. This work was funded by the Deutsche Forschungsgemeinschaft (DFG, German Research Foundation) under German’s Excellence Strategy—EXC-2123 QuantumFrontiers—390837967 and by DFG Grant No. ME 3648/5-1. H. A. F\"{u}rst was supported by the EMPIR project 18SIB05 "Robust Optical Clocks for International Timescales". 

\section*{References}
\bibliographystyle{iopart-num}
\bibliography{Paper}

\appendix
\section{Dynamical evolution of density matrix elements}
\label{sec:densitymatrix}
The equation for the evolution of the components of the density matrix of
vibrational states $n,m$ (see fig.~\ref{fig:quantummodel}(a)) takes the following form:
\begin{eqnarray}\label{eqn}
\frac{\partial \rho^{22}_{nm}}{\partial t} &=
&-(\gamma_3+\gamma_2)\rho^{22}_{nm} -i\omega_{osc}(n-m)\rho^{22}_{nm} \nonumber
\\&-&\frac{i}{2}\sum_\mu \left[\Omega^{(2)}_{n\mu} \rho^{12}_{\mu
m}-\rho^{21}_{n\mu}\Omega^{*\,(2)}_{\mu m}\right] \nonumber \\
\frac{\partial \rho^{11}_{nm}}{\partial t} &=
&-\gamma_1\rho^{11}_{nm}+{\hat
\gamma}_2 \{ \rho^{22} \}_{nm}  -i\omega_{osc}(n-m)\rho^{11}_{nm}\nonumber \\
&-&\frac{i}{2}\sum_\mu \left[\Omega^{*\, (2)}_{n\mu} \rho^{21}_{\mu
m}+\Omega^{(1)}_{n\mu} \rho^{01}_{\mu m}-\rho^{12}_{n\mu}\Omega^{(2)}_{\mu m}
-\rho^{10}_{n\mu}\Omega^{*\,(1)}_{\mu m}\right] \nonumber \\
\frac{\partial\rho^{00}_{nm}}{\partial t} &= &{\hat \gamma}_3 \{
\rho^{22} \}_{nm}+{\hat \gamma}_1 \{ \rho^{11} \}_{nm} -i\omega_{osc}(n-m)\rho^{00}_{nm}\nonumber \\
&-&\frac{i}{2}\sum_\mu \left[\Omega^{*\,(1)}_{n\mu} \rho^{10}_{\mu
m}-\rho^{01}_{n\mu}\Omega^{(1)}_{\mu m}\right] \nonumber \\
\frac{\partial \rho^{01}_{nm}}{\partial t} &=&
-\left(\gamma_1/2+i\delta_1\right)\rho^{01}_{nm} -i\omega_{osc}(n-m)\rho^{01}_{nm}\nonumber \\
&-&\frac{i}{2}\sum_\mu \left[\Omega^{*\,(1)}_{n\mu} \rho^{11}_{\mu
m}-\rho^{00}_{n\mu}\Omega^{*\,(1)}_{\mu
m}-\rho^{02}_{n\mu}\Omega^{(2)}_{\mu m}\right] \nonumber \\
\frac{\partial \rho^{10}_{nm}}{\partial t} &=&
-\left(\gamma_1/2-i\delta_1\right)\rho^{10}_{nm} -i\omega_{osc}(n-m)\rho^{10}_{nm}\nonumber \\
&-&\frac{i}{2}\sum_\mu \left[\Omega^{(1)}_{n\mu} \rho^{00}_{\mu
m}+\Omega^{*\,(2)}_{n\mu} \rho^{20}_{\mu m} -\rho^{11}_{n\mu}\Omega^{(1)}_{\mu
m}\right] \nonumber \\
\frac{\partial \rho^{12}_{nm}}{\partial t} &=&
-\left(\gamma_1/2+\gamma_2/2+\gamma_3/2+i\delta_2\right)\rho^{12}_{nm}\nonumber \\ &-&i\omega_{osc}(n-m)\rho^{12}_{nm} \nonumber \\
&-&\frac{i}{2}\sum_\mu \left[\Omega^{(1)}_{n\mu} \rho^{02}_{\mu
m}+\Omega^{*\,(2)}_{n\mu} \rho^{22}_{\mu m}
-\rho^{11}_{n\mu}\Omega^{*\,(2)}_{\mu
m}\right] \nonumber \\
\frac{\partial \rho^{21}_{nm}}{\partial t} &=&
-\left(\gamma_1/2+\gamma_2/2+\gamma_3/2-i\delta_2\right)\rho^{21}_{nm}  \nonumber \\
&-&i\omega_{osc}(n-m)\rho^{21}_{nm} \nonumber \\
&-&\frac{i}{2}\sum_\mu \left[\Omega^{(2)}_{n\mu} \rho^{11}_{\mu m}
-\rho^{20}_{n\mu}\Omega^{*\,(1)}_{\mu m}-\rho^{22}_{n\mu}\Omega^{(2)}_{\mu
m}\right] \nonumber \\
\frac{\partial \rho^{02}_{nm}}{\partial t} &=&
-\left(\gamma_2/2+\gamma_3/2+\gamma+i\delta_1+i\delta_2\right)\rho^{02}_{nm}\nonumber \\&-& i\omega_{osc}(n-m)\rho^{02}_{nm}\nonumber \\
&-&\frac{i}{2}\sum_\mu \left[\Omega^{*\,(1)}_{n\mu} \rho^{12}_{\mu
m}-\rho^{01}_{n\mu}\Omega^{*\,(2)}_{\mu
m}\right] \nonumber \\
\frac{\partial \rho^{20}_{nm}}{\partial t} &=&
-\left(\gamma_2/2+\gamma_3/2+\Gamma-i\delta_1-i\delta_2\right)\rho^{20}_{nm} \nonumber \\
&-& i\omega_{osc}(n-m)\rho^{20}_{nm}\nonumber \\
&-&\frac{i}{2}\sum_\mu \left[\Omega^{(2)}_{n\mu} \rho^{10}_{\mu m}
-\rho^{21}_{n\mu}\Omega^{(1)}_{\mu m}\right] \nonumber \\
\end{eqnarray}
where the superscripted indices $0,1,2$ of the density matrix elements denote the states $\ket{0}$, $\ket{1}$ and $\ket{2}$, respectively and the subscripted indices $n$, $m$ and $\mu$ are related to vibrational states. $\Gamma$ determines the decay of the coherence between the $\ket{2}$ and $\ket{0}$ states as a result of an uncorrelated phase of the laser fields ${\bf E}_1$ and ${\bf E}_2$.

The spontaneous relaxation operator ${\hat \gamma}\{\rho\}_{nm}$ determines the spontaneous decay rate $\gamma_1$, $\gamma_2$, and $\gamma_3$ (see fig.~\ref{fig:quantummodel} (a)), the decay of non-diagonal elements of the density matrix, as well as contributions from excited states to the ground state. These terms in (\ref{eqn}) have the form
\begin{equation}
{\hat \gamma}_i\{\rho^{jj}\}_{nm} = \sum_{\nu\mu} \Gamma^{(i)\, \nu\mu}_{nm}
\rho^{jj}_{\nu\mu}.
\end{equation}
The decay rates $\Gamma^{(i)\, \nu\mu}_{nm}$ for the dipole transitions $\ket{2}\to\ket{1}$ and $\ket{2}\to\ket{0}$ can be obtained from the general expression for the spontaneous relaxation operator taking into account recoil effects (see for example \cite{Prudnikov2007})
\begin{equation}\label{sp}
\Gamma^{(i)\, \nu\mu}_{nm} = \gamma_i  \int_{-1}^{+1} K^{(d)}(h) \left( C_{\nu
n}(\eta_i h)\right)^{+} C_{\mu m}(\eta_i h)\, dh,
\end{equation}
with $K^{(d)}(h)$ being the dipole pattern for the decay and $\eta_i$ are the
Lamb-Dicke parameters for the corresponding dipole transitions ($i = 2,3$ in fig.~\ref{fig:quantummodel} (a)). For the quadrupole transition the relaxation operator has a similar form to (\ref{sp}) with replacement of the dipole with the quadrupole pattern for the decay $K^{(q)}(h)$ and corresponding Lamb-Dicke parameter $\eta_1$~\cite{Kirpichnikova2019}.

\section{Off-resonant excitation via laser noise}
\label{sec:lasernoise}
\begin{figure}[h]
    \centering
    \includegraphics[width=0.7\textwidth]{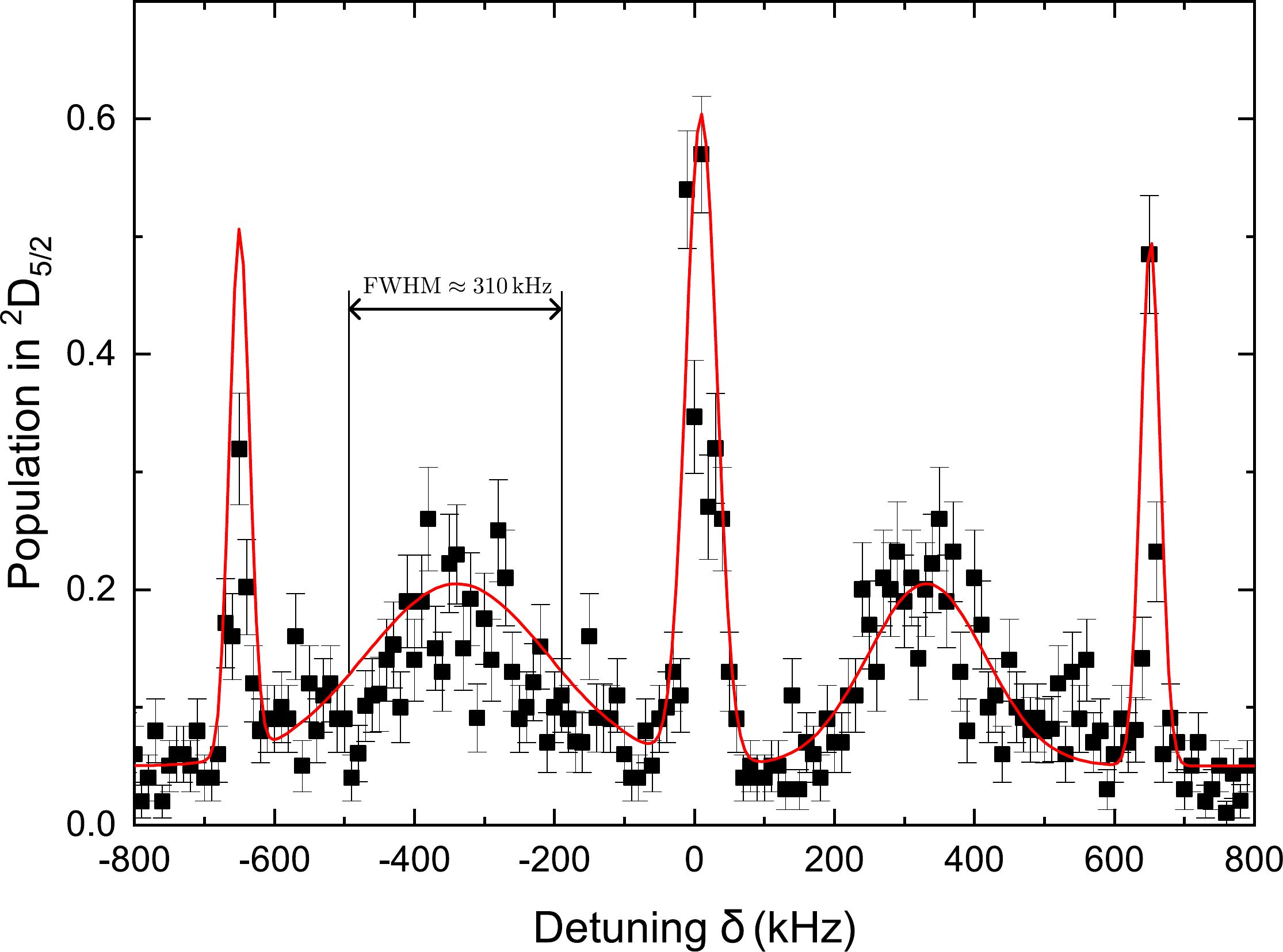}
    \caption{Frequency spectrum of the $^2$S$_{1/2}\to\,^2$D$_{5/2}$ transition near $\unit[411]{nm}$, relative to the carrier transition. For this measurement, the $^{172}$Yb$^+$ ion was confined with secular frequency $\omega_{\rm osc}=2\pi\times\,\unit[654(6)]{kHz}$ and interrogated with a pulse time of $t\simeq 7\cdot\tau_\pi$ to enhance the signature of laser noise between the sideband resonances. We fit Lorentzian functions to the carrier and the sidebands as a guide to the eye.   }
    \label{fig:noise}
\end{figure}

We attribute the faster cooling rate observed in our experiment at cooling laser detunings around $-1.5\,\omega_{\rm osc}$ (see fig.~\ref{fig:detuning}) to off-resonant excitation induced by laser noise. For verification, we studied the frequency spectrum of the $^2$S$_{1/2}\to\,^2$D$_{5/2}$ transition near $\unit[411]{nm}$ ranging from the first-order red sideband to the first-order blue sideband. Figure~\ref{fig:noise} shows such a frequency spectrum, recorded overdriven with interrogation time $t\simeq 7\cdot\tau_\pi$. The spectral features at $\pm\unit[350]{kHz}$ correspond to noise modulation of the laser light, most probably caused by the bandwidth of the locking electronics of the second harmonic generation cavity for $\unit[411]{nm}$. While the frequency detuning of the $\unit[411]{nm}$ cooling laser was set to $-1.5\,\omega_{\rm osc}$, with $\omega_{\rm osc}=2\pi\times\unit[565(5)]{kHz}$, both the first and second-order red sidebands were seperated by $\unit[283(5)]{kHz}$. With a FWHM of approx. $\unit[310]{kHz}$ of the noise spectral feature, a significant overlap to both red sidebands is given and most likely the reason for faster cooling observed in the experiment in this frequency range. 

\end{document}